# Navigating the Digital Chain in Concrete 3D Printing

Ali EL HAGE[1,2,3], Elodie PAQUET[1], Nordine LEKLOU[2], Thibault NEU[3], Philippe POULLAIN[2]

[1] Nantes Université, Ecole Centrale Nantes, CNRS, LS2N, UMR 6004, F-44000 Nantes, France
[2] Nantes Université, Ecole Centrale Nantes, CNRS, GeM, UMR 6183, F- 44600 Saint-Nazaire, France
[3] SEGULA Technologies, 1 Rue Charles Lindbergh, 44340 Bouguenais, France

`ali.el-hage@etu.univ-nantes.fr`

**Abstract.** The advancement of concrete 3D printing (C3DP) technology has revolutionized the construction industry, offering unique opportunities for innovation and efficiency. At the heart of this process lies a comprehensive digital chain that integrates various stages, from initial design to post-processing. This article provides an overview of this digital chain, explaining each crucial step. The chain begins with design, utilizing Design for Additive Manufacturing (DFAM) concept and parametric modeling to create optimized structures. Path generation follows, determining the precise toolpath for extruding concrete layers. Simulations, both numerical and analytical, ensure the design's integrity and feasibility. Several articles have addressed parametric modeling, process and numerical simulation, and the post-processing phase. However, none has proposed an updated methodology for the workflow. This study aims to propose a robust digital chain for C3DP technology, using one platform (3Dexperience) and seamless data transfer between applications. These steps provide insights into the structural performance of printed components, enabling necessary adjustments and optimizations. In essence, the digital chain coordinates a seamless workflow that transforms digital designs into concrete structures, unlocking the full potential of C3DP and paving the way for innovative and efficient construction.

## 1 Introduction

C3DP is an example of how digital technologies and conventional building techniques have come together to create a new era of efficiency and innovation. This innovative method of construction promises exceptional design freedom, quick prototyping, and environmentally friendly building techniques[1]. A robust digital chain that leads the process from conception to implementation is at the center of this significant process. In this study, the details of the digital chain for C3DP are explored, shedding light on the key stages and methodologies driving this technology forward. From the initial stages of design and path generation, simulation, printing, and post-processing, each step in the digital chain plays a crucial role in shaping the final outcome.

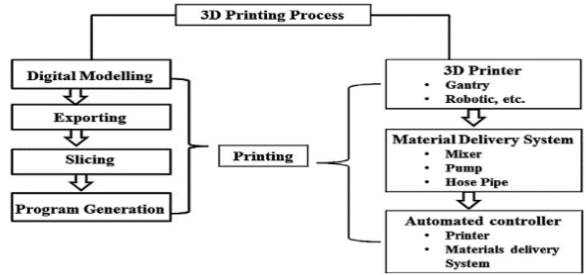

**Fig. 1** The process of 3D printing for the construction industry **[2]**

Fig. 1 illustrates the typical workflow of C3DP within the construction industry, presenting a comprehensive overview of both software and hardware segments involved in the process [2]. In the software aspect, users employ 3D modeling software to generate detailed models of desired objects. These models are then exported in stereolithography (.stl) format to slicing software. In slicing software, layer dimensions are specified, allowing the model to be segmented into printable layers. The sliced model then undergoes the generation of a program file, which serves as precise instructions for the 3D printer to execute the fabrication process. The digital chain's role in streamlining the printing process of C3DP within the construction industry underscores the importance of having a robust framework in place. A well-structured digital chain ensures seamless coordination between various software and hardware components, minimizing the risk of failure during printing and optimizing the utilization of AM technology.

## 2 Preprocessing

### 2.1 Design and preparation

CAD softwares were initially developed for traditional manufacturing processes like milling and continues to rely on the combination of low-level geometric primitives using sketch-based modeling[3]. Sketch-based modeling is commonly used to create basic parts, often without precise curves, as curves are generated randomly. This method is followed by the slicing process. However, a drawback of sketch-based modeling is the lack of complete or significant control over the toolpath during the concrete 3D printing process. DfAM focuses on customizing product designs to suit the capabilities and limitations of AM technology. In the context of C3DP, DfAM allows for the optimization of structures, improving material efficiency, accommodating geometric complexity, and enhancing buildability. Topology optimization, a DfAM computational method, optimizes material distribution for specific performance goals while minimizing usage. It's vital for crafting lightweight structures suited for AM, reallocating material for efficiency, ideal for C3DP. [4].
Both parametric modeling and generative design serve as essential tools within the realm of computational design[5]. Parametric modeling involves creating design models with adjustable parameters, enabling the exploration of various iterations and configurations efficiently. This approach facilitates the rapid iteration of designs and allows for easy adaptation to changing requirements or constraints. Mathematical equations are often used in parametric modeling to define geometric surfaces, curves, and forms. A circle, for instance, can be characterized by a single equation that expresses its radius, but more complex forms can need multiple equations to fully capture its form. The model form can be altered by modifying these equations or their parameters, giving flexibility and control. In Fig. 2, the motion is described by a parametric equation in the x-y plane. Where $R_C$ denotes the radius from the origin, determining the starting distance of the point from the center. The sinusoidal variation in the radius of the circular motion varies with the amplitude ($a$) of oscillation and ($n$) which represents the frequency of the oscillation. Moreover, the application of knitting and woven patterns such as the one shown in Fig. 3 enhances architectural design by enabling precise control over material distribution, thereby facilitating the creation of intricate patterns and textures in structures. This technique's precision offers flexibility, allowing for the execution of complex designs with minimal user intervention. Additionally, it provides versatility in forming various stitches and patterns, which facilitates varying the porosity and density of the structure.

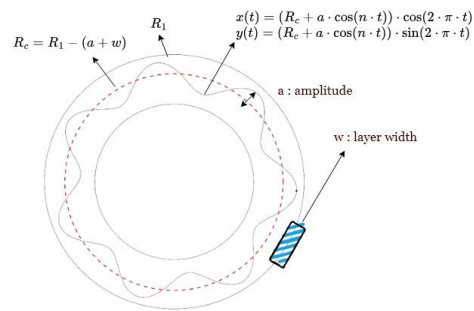

**Fig. 2** Parametric sinusoidal oscillation within a circular motion

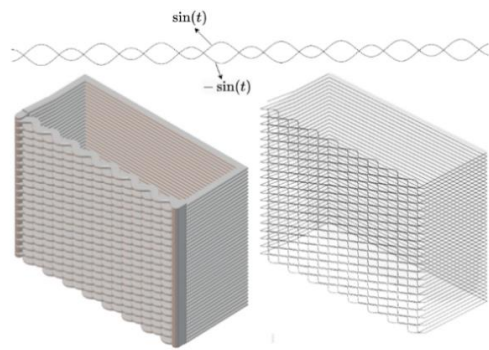

**Fig. 3** Parametric woven pattern

Slicing is a preparatory stage preceding the manufacturing process for a CAD model. Currently, the predominant approach in AM is the planar slicing method. However, this technique has drawbacks, notably excessive material deposition at transition points leading to over-extrusion, causing issues such as poor surface finish, and dimensional inaccuracies. Overcoming such problems can be achieved by employing non-planar slicing strategies. One such method is helical slicing, which eliminates discontinuous extrusion tool paths and creates a single continuous extrusion path with a single start and end point. By doing so, defects at the start and end points of extrusion layers are significantly reduced, effectively eliminating seams in the manufactured part. Additionally, this method reduces non-extrusion movements, further enhancing the quality of the final product[6].

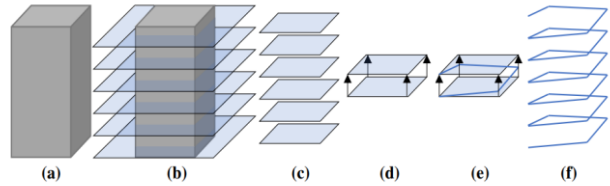

**Fig. 4** The helical slicing method [6]

Instability is primarily correlated with the model's geometric complexity such as curved surface, twisted, and overhang structures. For instance, in a twisted structure like that illustrated in **Fig. 5**, transitioning from a rectangular to a circular form often leads to misaligned layers and instability. Implementing a helical path with precise motion during printing is essential to restore stability and prevent such misalignment issues.

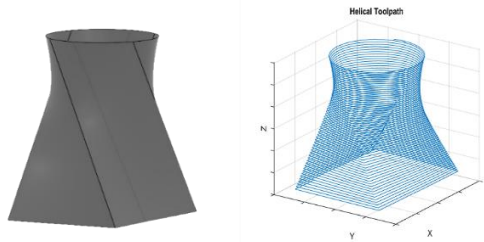

**Fig. 5** Twisted model helical path

## 3 Simulation

### 3.1 Process Simulation

Process simulation is crucial for additive AM processes as it enables accuracy, effectiveness, and optimization throughout the entire workflow [7]. This technique employs virtual modeling to precisely plan the actions of robotic machines, ensuring precise material deposition, smooth execution of toolpath, and careful positioning of the robot on the construction site. Toolpath simulation meticulously coordinates the movement of the robot as it traverses through the complex layers of material deposition. Moreover, it allows placement optimization, and collision detection algorithms to safeguard equipment and surroundings.

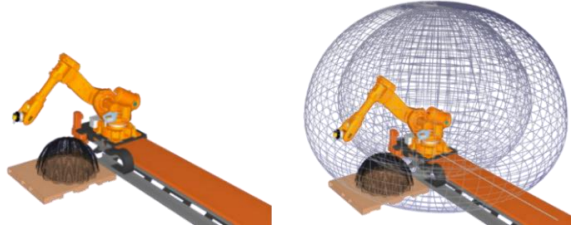

**Fig. 6** ABB IRB 6640 workspace

Material deposition simulation focuses solely on the process of depositing the materials layer by layer, without considering mechanical properties or behavior. In this simulation, the user has control over how each layer will be deposited to ensure the uniformity and consistency of each printed layer. parameters such as nozzle speed, material flow rate, and layer thickness can be adjusted to achieve the desired characteristics for each layer. A critical consideration is ensuring that within the same layer, there are no collisions in the material. This means avoiding over-deposition caused by errors in the toolpath simulation for complex geometries (Fig. 7).

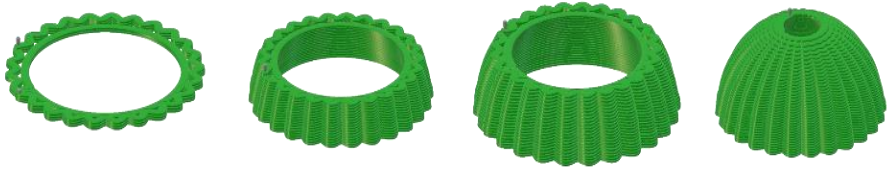

**Fig. 7** Material deposition simulation

### 3.2 Numerical Simulation

Simulating the AM process is invaluable in preventing structural failures by identifying geometrical issues, ensuring structural safety, understanding the underlying physics, and optimizing both the process and structural design. The main failure mechanisms in C3DP are plastic collapse and elastic buckling [8]. Different researchers have explored various methods of simulating the printing process [9], [10], [11]. The numerical simulation employed in this proposed digital chain is discussed in [9], where the Mohr-Coulomb criterion is applied to analyze the failure of the printed structure using Eq. (1). Additionally, apart from the necessary printing parameters (printing speed, layer dimensions, etc.), the time-dependent equation for the cohesion (C) property is required. Here φ signifies the angle of internal friction resulting from the frictional resistance and interlocking among internal particles. Both C and φ may vary over time. The shear yield stress and acting normal stress are denoted by $\tau_y$ and $\sigma_n$, respectively.

$$\tau_y = C(t) + \sigma_n \cdot tan\big(\varphi(t)\big) \tag{1}$$

Since printing failures are also stability-driven (buckling), the use of the stress-strain relation before failure is important. Therefore, the stiffness of the material, represented by the Young's modulus E(t), and the Poisson ratio ν(t) are essential parameters. These parameters help model the object's response before yielding occurs. Using an unconfined compression test, both parameters can be measured [9]. The execution of this numerical simulation requires both the dilatancy angle and the density of the material used. In this part of the digital chain, numerical simulations are conducted using 3Dexperience SIMULIA (ABAQUS). Following the acquisition of the toolpath trajectory detailed in the preceding section 3.1, the trajectory is then transformed into an event series. This event series facilitates synchronization between element activation and the toolpath trajectory, enabling activation of elements and initiation of material properties behavior following predetermined equations when material deposition occurs. When it comes to predicting failure, numerical simulation of the printing process primarily focuses on complex geometries rather than monolithic structures. C3DP offers a distinct advantage in its ability to fabricate overhang structures. However, the feasibility of printing such structures depends on various factors, including material properties, toolpath design, and other printing parameters and configurations such as printing orientation and layer dimensions. To enhance the numerical model and improve the visualization of failure modes, numerical damping was incorporated. Fig. 8 presents two examples of numerical simulations.

To enhance the numerical model and facilitate a better understanding of the failure modes, numerical damping was introduced. Fig. 8 presents two examples of numerical simulations. The first illustrates the failure and deformation measurements of a dome structure, while the second shows the failure and structural changes of a cylindrical form, and the stress distribution based on the von Mises yield criterion.

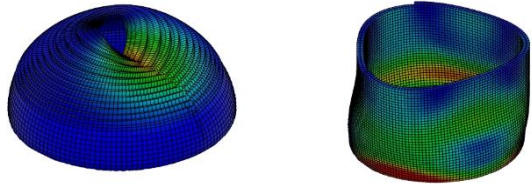

**Fig. 8** Numerical Simulation with different configurations

## 4 Post Processing

Ensuring the quality, integrity, and compliance of 3D printed concrete structures with established standards and specifications is crucial for successful technology implementation. Inspection and quality control traditionally involve manual measurement of structural and layer dimensions. In this context, 3D scanning technology is proposed to automate and enhance the precision of inspection processes for printed structures. This advancement enables detailed analysis of deformations and defects arising from tool-path or material properties (**Fig. 9**). The scanned model can be compared to the original CAD model, allowing for a direct assessment of dimensional accuracy and alignment with design specifications. Moreover, data from 3D scans can undergo processing using advanced image processing algorithms to detect and characterize deformations, cracks, or other irregularities within printed components[12]. These capabilities facilitate the identification of even subtle deviations from the intended geometry, supporting prompt corrective actions to maintain structural integrity and performance standards.

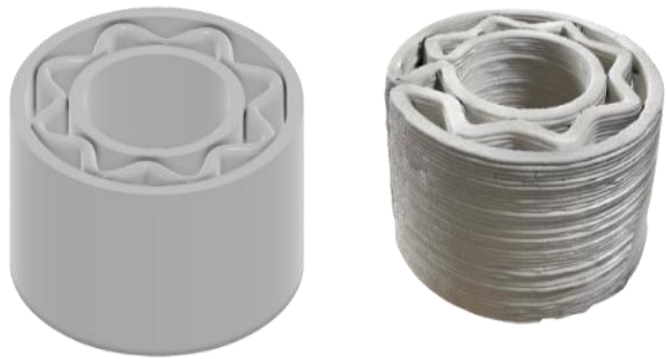

**Fig. 9** Designed and printed structure

## 5 Conclusion

The effective integration of a comprehensive digital chain is important for the advancement of C3DP within the construction industry. From initial design considerations to meticulous post-processing inspections, each stage in the digital chain plays a pivotal role in ensuring the efficiency, quality, and compliance of 3D printed structures. Embracing methodologies like Design for Additive Manufacturing (DfAM) and parametric modeling, alongside advanced simulation techniques, allows for the optimization of designs and printing processes. However, the future perspective of C3DP extends towards establishing an even more robust digital chain, facilitating seamless transitions between each phase. This involves the integration of automated analysis tools for geometry and texture measurements using sophisticated image processing techniques during post-processing. Leveraging insights gained from these analyses not only identifies areas for enhancement but also enables the refinement of printing parameters to elevate quality and consistency. Additionally, we should also benefit from experimental results to enhance our numerical database, ensuring the accuracy and reliability of simulations. As the technology continues to evolve, optimizing the digital chain will be instrumental in unlocking the full potential of C3DP, empowering the construction industry with unprecedented levels of innovation and efficiency.